\documentclass[runningheads]{llncs}

\usepackage{eccv}

\usepackage{eccvabbrv}

\usepackage{graphicx}
\usepackage{booktabs}
\usepackage{multirow}
\usepackage{floatrow}
\usepackage{caption}
\usepackage{tabularray}

\usepackage[accsupp]{axessibility}  % Improves PDF readability for those with disabilities.

\usepackage{hyperref}

\usepackage{orcidlink}

\begin{document}

\title{GeoScaler: Geometry and Rendering-Aware Downsampling of 3D Mesh Textures} 

\titlerunning{Abbreviated paper title}

\author{Sai Karthikey Pentapati\inst{1,2}\and
Anshul Rai\inst{2}\and
Arkady Ten\inst{2}\and
Chaitanya Atluru\inst{2}\and
Alan Bovik\inst{1}}

\authorrunning{F.~Author et al.}

\institute{The University of Texas at Austin \and
Dolby Laboratories Inc., Sunnyvale, CA}
\maketitle

\begin{figure}
    \centering
    \includegraphics[width=12cm]{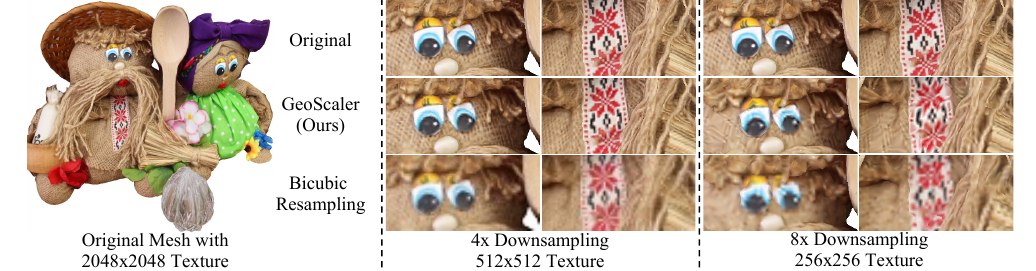}
    \captionof{figure}{
    \textbf{Qualitative comparisons of downsampling.} GeoScalar can downsample texture maps of 3D meshes while providing significantly improved quality of rendered images as compared to existing methods such as bicubic resampling. The results shown here were computed on the \textit{clothdolls} mesh from our 3DSet5 mesh dataset.}
    \label{fig:title}
\end{figure}
\vspace{-25px}
\begin{abstract}
  High-resolution texture maps are necessary for representing real-world objects accurately with 3D meshes. 
The large sizes of textures can bottleneck the real-time rendering of high-quality virtual 3D scenes on devices having low computational budgets and limited memory.
Downsampling the texture maps directly addresses the issue, albeit at the cost of visual fidelity.
Traditionally, downsampling of texture maps is performed using methods like bicubic interpolation and the Lanczos algorithm. 
These methods ignore the geometric layout of the mesh and its UV parametrization and also do not account for the rendering process used to obtain the final visualization that the users will experience.
Towards filling these gaps, we introduce GeoScaler, which is a method of downsampling texture maps of 3D meshes while incorporating geometric cues, and maximizing the visual fidelity of the rendered views of the textured meshes.
We show that the textures generated by GeoScaler deliver significantly better quality rendered images compared to those generated by traditional downsampling methods.
\end{abstract}

\section{Introduction}
\label{sec:intro}
Textured meshes have proven to provide efficient representations of 3D objects in terms of representation density, quantization, and ease of rendering.
For example, point clouds are often too sparse, voxel grids suffer from memory inefficiencies, and volumetric representations are difficult to render. 
Textured meshes are less affected by these limitations, and because of their versatility, they are ubiquitous in real-time computer graphics applications like video games, and most graphical hardware products are designed and manufactured to accelerate their rendering.
However, real-time, high-quality graphics applications are still difficult to realize on low-compute devices such as virtual reality headsets and smartphones.
This has forced developers and graphic artists to either build meshes containing smaller texture maps or downsample their existing high-resolution texture maps, leading to a compromise in visual quality. 

Contemporaneous methods such as bilinear, bicubic, and Lanczos \cite{lanczos1988applied} resampling are very commonly used for downsampling textures. These methods do not adapt to the data geometry, as they were designed for planar two-dimensional images.
Planar images differ from texture maps in two key ways, making standard resamplers inefficient for processing 3D mesh textures.

\begin{figure}
    \centering
    \includegraphics[width=12cm]{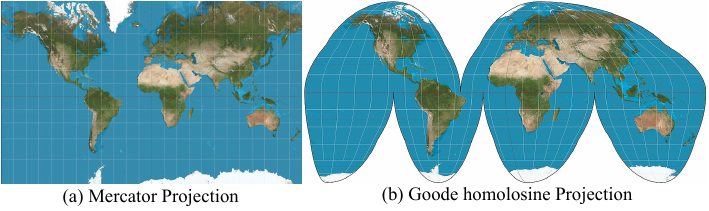}
    \caption{\textbf{The trade-off between warping and discontinuities when creating UV parametrization.} (a) Regions like Greenland and Antarctica which lie near the poles consume disproportionately large areas on Mercator projection. (b) Goode homolosine projection reduces warping artifacts but introduces many discontinuities.}
    \label{fig:projection}
\end{figure}

Unlike regular 2D images, texture maps contain discontinuities and warping artifacts that arise during UV mapping, which is the process of projecting the 3D surface of a mesh onto a 2D plane.
As it is often impossible to flatten a 3D surface onto a two-dimensional plane without causing major distortions, it is required to slice the 3D surfaces into several patches, and then flatten each patch separately.
\cref{fig:projection} illustrates the trade-off between these two types of artifacts using different projections of the spherical surface of the Earth onto a 2D map.
The Mercator projection (\cref{fig:projection}a) of the world map significantly distorts areas near the two poles by expanding their surface areas more than on regions closer to the equator. 
Conversely, the Goode homolosine projection \cite{Goode1925THEHP} (\cref{fig:projection}b) reduces the distortions but introduces significant discontinuities.

The aforementioned resampling methods operate by applying fixed-size symmetric kernels on images to estimate the resampled values.
Applying simple interpolation kernels at the discontinuities on texture maps leads to blur and/or boundary displacements.
Due to the localized nature of the applied kernels, true neighbors of a pixel sitting on a discontinuity in a texture map may not be involved while filtering that pixel. %?%
% This can be observed in the mismatch between the true effect of the kernel (blue patch) and the desired effect (green patch) in \cref{fig:projection}c, where applying the kernel on a boundary of the world map generated using the Goode homolosine projection.
Another problem arises from the symmetric nature of standard interpolation kernels, whereby pixels in all directions are weighted similarly.
However, the assumption of symmetric distances is not generally true for texture maps.
Textures become warped when the 3D surfaces they are painted on are flattened onto 2D planes, as the UV parametrization is always non-isometric for non-planar surfaces: 
areas of the triangles on the mesh in 3D are non-uniformly modified when mapped to their 2D representation on the UV plane.
% For instance, the mismatch between the effect of applying a square kernel on the highly warped regions of the Mercator projection map (red patch) and the desired effect (green patch) can be noticed in \cref{fig:projection}c
Ideally, the downsampling algorithm should account for warping by flexibly weighing the pixels during the interpolation

Furthermore, the synthesis of the UV parametrization does not take into account the content of the texture itself. 
This can lead to small areas on the 2D texture maps being assigned to regions with intricate color/attribute texture detail. 
The quality of the rendered images could be better if large but less-detailed triangles were assigned smaller areas and small but highly-detailed triangles were assigned larger areas.

Finally, texture maps are viewed by users in the form of rendered 2D views. 
Thus, it is crucial to ensure that the downsampling process is designed to optimize the perceived visual quality of the rendered content, instead of prioritizing data fidelity of the texture on the UV plane.

These observations demonstrate the interdependence between the geometry of a mesh, its UV parametrization, and its texture map. 
In our view, this motivates the need for the development of downsampling techniques for texture maps that account for these characteristics of texture maps, as well as the rendering process used to generate the final view of a mesh.

\begin{figure*}
    \centering
    \includegraphics[width=12cm]{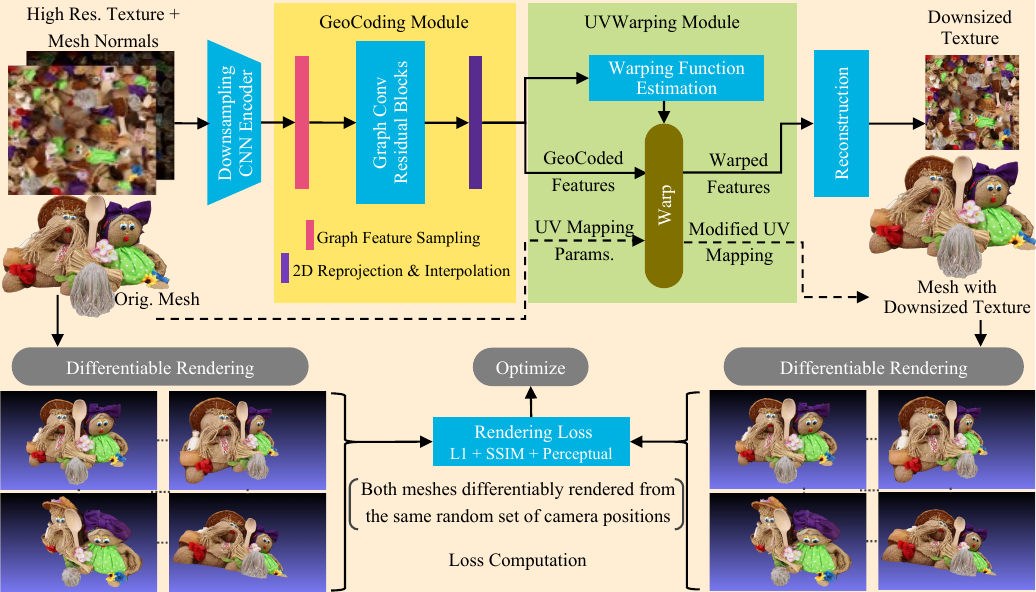}
    \caption{\textbf{Overview of the GeoScaler architecture.} For a given input textured 3D mesh, an optimal texture downsampling function is learned which incorporates information of the geometry of the mesh and its UV parametrization. The model parameters are iteratively updated via differentiable rendering by computing the loss between views rendered from the original texture of the mesh versus the downsampled texture.}
    \label{fig:overview}
\end{figure*}

Towards this end, we have created GeoScaler, which is a texture map downsampling method that addresses the drawbacks of existing methods. 
An overview of our method is shown in \cref{fig:overview}.
GeoScaler is a deep neural network (DNN) based downsampling model that is optimized to maximize the perceptual quality of images rendered using the downsampled texture map of the mesh.

The design of GeoScaler is based on three key concepts: %%%%%%
\begin{enumerate}
    \item Rendering loss is calculated between views rendered from the mesh with its original texture map, and the mesh with a downsampled texture map generated by the DNN. This loss is used to optimize the network.
    \item The GeoCoding module exploits the geometric layout and the UV parametrization of the mesh to counteract the effects of warping distortions and discontinuities in the texture map 
    \item The UVWarper module computes a warping function that adjusts the UV parameters of all the vertices of the mesh to obtain a more efficient layout of the texture based jointly on a mesh's geometry and RGB texture.
\end{enumerate}
In the experiments, we demonstrate that GeoScaler significantly outperforms existing downsampling methods on a wide variety of meshes by providing up to 3dB and 2dB improvement for 8x and 4x downsampling of textures respectively.

\section{Related Works}
\label{sec:related}
\subsection{Generation of Textured Meshes}
In most interactive 3D graphics applications, such as video games and virtual/augmented reality applications, the majority of graphics assets are depicted using textured meshes.
These meshes are usually designed by graphics artists and creatives, who build complex objects by blending primitive 3D shapes or sculpting complex geometric surfaces. 

However, there has also been a significant increase in 3D content captured from real-world scenes in recent years driven by rapid improvements in 3D reconstruction techniques such as structure from motion (SfM) \cite{schoenberger2016sfm,wei2020deepsfm,xu2023unifying} and implicit representations like \cite{mildenhall2020nerf,yu2022plenoxels,sitzmann2019siren}.
While methods exist that can directly generate meshes from input views like \cite{goel2021differentiable,chen2022mobilenerf}, most of these methods produce 3D reconstructions in the form of point clouds, voxel grids, or volumetric fields which have to be converted into textured meshes using algorithms like Poisson mapping \cite{poiss} and marching cubes \cite{LorensenC87}.

When generating a texture map from textured 3D mesh surfaces, a UV parametrization process is performed which involves unwrapping the 3D model by generating a mapping from each point on its surface onto the two-dimensional UV plane.
Popular UV mapping techniques include \cite{10.1145/3130800.3130845, 10.1111:1467-8659.00580, uvm, 10.1145/3272127.3275042}.
To reduce distortions that arise during flattening from 3D to 2D, it is common to partition each mesh into multiple patches, whereby the flattening and UV parametrization are done independently on each patch.
Multiple ways of decomposing meshes into patches have been proposed  \cite{1183787,liu2022dawand,1335424}. 

While these methods minimize seams and geometric distortion within texture maps, they do not pay heed to the RGB content of textures.
Additionally, noise arising during the 3D reconstruction of geometry often causes unnatural texture seams, inefficient utilization of space on the UV plane, and sub-optimal calculation of the UV mapping.
These drawbacks can lead to inefficient usage of space on texture maps and thus also memory usage while rendering and reduced perceptual quality of the rendered images.
These problems are generally exacerbated when the resolution of the texture map is reduced.

\subsection{Efficient and Alternate Texture Representation}
While there has not been much focus on the downsampling of 2D texture maps, there has been significant research on compressing texture maps, rendering from compressed textures, and deriving alternate efficient representations of surface color/attributes. 
Space-filling 3D textures (or solid textures) were proposed in \cite{10.1145/325334.325247, 10.1145/325165.325247} to counteract the detrimental effects of texture seams and warping.
To combat high memory usage while rendering, the authors of \cite{beers1996rendering,  10.1145/2980179.2982439, ning1993fast, inproceedings} proposed algorithms to build compressed representations of texture maps and directly render from them.
The authors of both \cite{nasiri:hal-02010328} and \cite{ 10.2312:EGGH:HPG12:105-114} both aim to modify standard image compression algorithms by introducing geometry-aware methods for intracoding. 
All of these methods can be applied along with our method to gain the benefits of both efficient compression and downsampling.

Mip ("multum in parvo") Mapping is commonly applied on texture maps to reduce aliasing when rendering meshes from a distance.
This process also involves sub-sampling textures at different scales, using methods like bilinear or trilinear filtering.
While some effort has been made \cite{765326, 10.1145/1077534.1077564} to optimize mipmap generation, these techniques do not facilitate the generation high quality renders with downsampled textures at the primary viewing distance. 

\subsection{Deep Learning on Meshes}
In recent years, many deep learning-based techniques have been proposed for applications related to textured meshes including 3D object recognition \cite{LIANG2022109500, lahav2020meshwalker}, detection \cite{brazil2022omni3d, Kulkarni-2019-121566}, and segmentation \cite{10.1145/3306346.3322959, Kalogerakis:2010:labelMeshes}. 
These efforts have built upon modifications of graph convolutions \cite{kipf2016semi} such as \cite{10.1145/3306346.3322959, NEURIPS2020_0a656cc1, 10.1145/3506694}.
While graph convolutions are suitable for processing geometry and connectivity of meshes and ordinary Cartesian convolutions are appropriate on 2D texture maps, there are no frameworks for processing them jointly.

A variety of deep models have been proposed for generating 3D meshes from multiple \cite{goel2021differentiable, usl2022} or single images \cite{Tian2022Recovering3H, wang2018pixel2mesh}. 
A key ingredient of these generative models is differentiable rendering, which allows parameters of a 3D scene to be updated using measurements made on rendered 2D views.
A plethora of differentiable rendering algorithms have been proposed \cite{Kato_2018_CVPR, loper2014opendr} that implement differentiable rasterization \cite{liu2019softras, liu2020general} and path-tracing \cite{8953665, Li:2018:DMC}.

% \subsection{2D Image Downsampling}
% In addition to traditional methods like bicubic and Lanczos resampling \cite{lanczos1988applied}, other methods to downsample images have also been proposed.
% These include per-image optimization methods such as \cite{sun2020learned} and \cref{kopf2013content} which try to match the super-resolved versions of downsampled images with the ground truth and all-purpose methods like \cite{Oztireli15Perceptual}. These methods have not been specifically adapted to textures of 3D meshes and lack awareness of the geometry of the rendering space.

\section{Method}
\label{sec:method}

\subsection{Overview}

GeoScaler takes textured 3D meshes as input, learns an optimal texture downsampling function, and generates optimally downsampled textures.
Unlike traditional methods, where texture downsampling is performed without knowledge of the mesh geometry and its UV parametrization (as shown in  \cref{eqn:traditional}), GeoScaler exploits these properties in addition to knowledge of the rendering process to generate an optimally downsampled texture map that maximizes the visual fidelity of the rendered views. 
GeoScaler can be expressed as in \cref{eqn:geoscaler}, where \(\mathbf{T}\) is the original texture map, \(\mathbf{T}_{\downarrow s}\) is the texture map downsampled by a factor of \(s\), \(\mathbf{M}\) is the mesh, and \(r(\cdot)\) is the rendering function used to transform the 3D mesh into a 2D view. The mesh \(\textbf{M}\) includes the vertex positions \(\mathbf{V}\), edge connectivity \(\mathbf{E}\), and the UV coordinates of all vertices \(\mathbf{U}\). The set $\mathbf{V}=\{v_j\}_{j=1}^N, v_j\in\mathbb{R}^3$ contains the position of $N$ vertices in the 3D space, $\mathbf{U}=\{(u_j^x, u_j^y)\}_{j=1}^N, u_j^x, u_j^y\in\mathbb{R}$ are the UV parameters corresponding to each vertex and $E$ is the adjacency list of connections between vertices:
\begin{equation} \label{eqn:traditional}
    \mathbf{T}_{\downarrow s} = \text{DownSampler}_s(\mathbf{T})
\end{equation}
\begin{equation} \label{eqn:geoscaler}  
    \mathbf{T}_{\downarrow s} = \text{GeoScaler}_s(\mathbf{M}(\mathbf{T}, \mathbf{V}, \mathbf{E}, \mathbf{U}), r(\cdot))
\end{equation}

GeoScaler computes an optimal downsampling function by training a mesh-specific DNN to downsample its texture map.
For every input textured mesh, GeoScaler optimizes the parameters of a deep neural network that downsamples the textures of the mesh in an iterative self-supervised manner, by minimizing the differences between images rendered from the original mesh+ texture and the mesh with downsampled textures.
This process can be formulated as shown below in \cref{eqn:overview} and \cref{eqn:optim}.

\begin{equation} \label{eqn:overview}
\mathbf{T}_{\downarrow s}, \mathbf{U'} = \text{DNN}(\mathbf{M}(\mathbf{T}, \mathbf{V}, \mathbf{E}, \mathbf{U}), \theta_{opt})
\end{equation}
where
\begin{equation} \label{eqn:optim}
\begin{gathered}
    \theta_{opt} = \arg\min_{\theta} \text{RenderLoss}(\mathbf{M}(\mathbf{T}_{\downarrow s}, \mathbf{V}, \mathbf{E}, \mathbf{U'}), \\ \mathbf{M}(\mathbf{T}, \mathbf{V}, \mathbf{E}, \mathbf{U})).
\end{gathered}
\end{equation}

\subsection{Model Architecture}
Our proposed deep network architecture enables leveraging cues from the geometric layout of the mesh and its UV parametrization to improve the downsampling performance.
Given an input mesh \(\mathbf{M}(\mathbf{T}, \mathbf{V}, \mathbf{E}, \mathbf{U})\)  with a texture size \(H\times W\), the first step involves projecting the three-dimensional geometric properties of the mesh, specifically the vertex positions and normals, onto a two-dimensional map that aligns with the texture map. 
This can be done by utilizing barycentric coordinates to first smoothly interpolate the geometric features over the surface of the mesh, then using the UV parametrization to flatten them onto a two-dimensional feature map.
Since the same UV parametrization is used to project textures onto the surface of the mesh, the texture map will align with the geometric feature maps.
The conversion of geometric information to the 2D domain allows it to be filtered with ease via regular convolutional layers.
This idea of converting geometric information to texture-aligned 2D features has been previously successfully applied for 3D geometry editing \cite{DBLP:journals/tog/AlliezMD02}.
These feature maps, which contain rich information regarding the three-dimensional properties of the mesh, are concatenated with the texture map before being input to the DNN. 
This simple step allows the DNN to access the geometric properties of the mesh when determining the optimal parameters for downsampling in a geometry-aware manner.

The input texture map and the geometric features are then processed via an encoder-like module comprising of convolution and pooling layers which extract a set of deep features \((F_e)\) at the desired reduced resolution of \(H/s\times W/s\) and $C_1$ channels.
This is followed by the GeoCoding module, which explicitly filters the deep features per the connectivity and layout of the mesh.
Before reconstructing an RGB texture map, the UVWarper module estimates the warping function to be applied to the texture map and the UV parameters of the mesh vertices.
The warping aims to make the texture layout and the UV mapping of the mesh more efficient.
Each of these modules is explained in detail in the following sections.

\subsection{GeoCoding Module}
\label{subsec:gcm}
\begin{figure*}
    \centering
    \includegraphics[width=12cm]{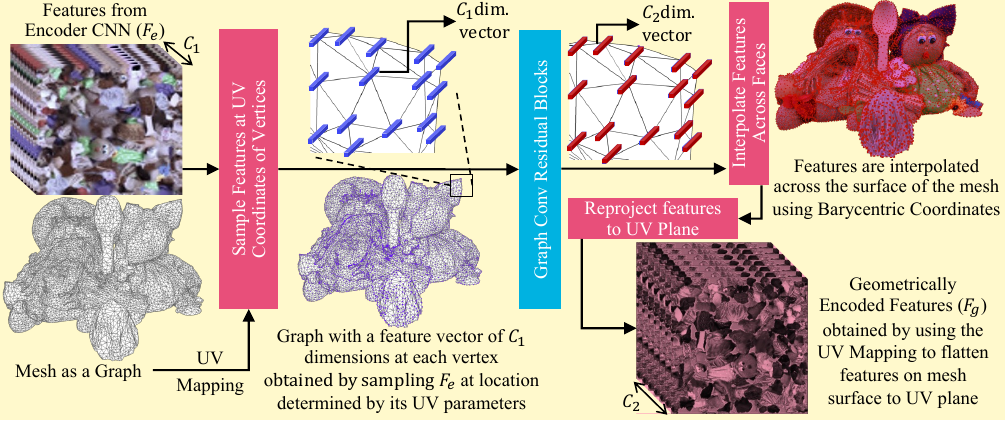}
    \caption{\textbf{Overview of the GeoCoding module.} Features obtained from the texture encoder are mapped to vertices on the surface of the mesh using the UV parametrization of the mesh. Performing graph convolution operations on these mapped features allows combining geometric information from the mesh with texture information, which reduces artifacts at texture map discontinuities. The output features are then interpolated using Barycentric coordinates and reprojected back to the UV plane.}
    \label{fig:geocoder}
\end{figure*}

Due to the limited receptive field of convolution operations, pixels on the boundaries of a texture map are not filtered with their true neighbors on the 3D surface.
The GeoCoding (short for "geometric encoding") module aims to address artifacts that arise due to these discontinuities of the texture map.
The structure of the GeoCoding module is depicted in \cref{fig:geocoder}.

The first step involves using the UV parametrization $(\mathbf{U})$ to sample a $C_1$-dimensional feature embedding ($f_i$) at the location of each of the \(N\) mesh vertices from $F_e$, the $C_1\times H/s\times W/s$ dimensional feature map produced by the CNN-based encoder.
\cref{eqn:sampler} denotes this process:
\begin{equation} \label{eqn:sampler}
    \begin{gathered}
        u_i^x, u_i^y = \textbf{U}(v_i), i=1...N \\
        f_i = F_e[u_i^x, u_i^y]
    \end{gathered}
\end{equation}
If a vertex appears more than once in the UV parametrization, we average the samples from each position.
After sampling, we obtain a graph with $N$ vertices and a feature embedding  $(f_i \in\mathbb{R}^{C_1})$ associated with each vertex.
Unlike the 2D feature map which suffers from discontinuities due to the UV mapping, these vertex-wise feature embeddings $(f_i)$, albeit sparse, lie adjacent to their true neighbors on the 3D surface and are connected to them via the edges of the mesh.

The vertex-wise feature embeddings are filtered using residual blocks built using simple graph convolution operations as shown in  \cref{eqn:gconv}  which is borrowed from \cite{10.1609/aaai.v33i01.33014602}.
\begin{equation} \label{eqn:gconv}
    \begin{gathered}
        f'_i = \theta_1f_i + \theta_2\sum_{j\in N(i)}e_{ij}f_j,
    \end{gathered}
\end{equation}
where \(N(i)\) is the set of neighbors of \(i\), \(e_{ij}\) is the weight between nodes \(i\) and \(j\), and \(\theta_1\) and \(\theta_2\) are learned parameters.
We designate the edge weights \(e_{ij}\) to equal the Euclidean distance between the mesh vertices.
Performing the convolutions in the geometric domain ensures that features that are proximal in the 3D domain on the surface of the mesh, but lying on distant seams of texture maps, are appropriately fused.
The sequence of graph convolutions outputs a set of vertex-wise feature vectors of length $C_2$.

To reconstruct a two-dimensional map from geometrically filtered features, we first smoothly interpolate the features sampled at each vertex across the faces using barycentric coordinates.
Next, the UV parametrization is again used to project features on the mesh surface onto the 2D UV plane.
The output of the GeoCoding module, a two-dimensional feature map $(F_g)$ of size \(C_2\times H/s\times W/s\) is generated.
The obtained feature set $F_g$ is compensated for geometric discontinuities and distortions present in the texture map.
If vertices are extremely sparse, and the sampled vectors at the vertices fail to retain sufficient information,  $F_g$ is added with a skip residual from $F_e$. 
The need for this skip residual is demonstrated more thoroughly in supplemental section \cref{subsec:s1}

\subsection{UVWarper Module}
Before reconstructing the downsampled RGB texture map, a warping function $f_{warp}: \mathbb{R}^2\rightarrow\mathbb{R}^2$ is estimated that aims to improve the UV parametrization.
This warping aims to optimize the layout of the texture map by assigning larger areas to more intricate textures.
We estimate $f_{warp}$ by deriving a sub-pixel offset at each pixel on the texture map from a slice of the geometrically encoded features $F_g$.
The predicted offset is then applied to the remaining slice of $F_g$, to obtain a texture map features $F_w$ having an optimized layout.
A new UV parametrization $\textbf{U'} = f_{warp}(\textbf{U})$ is also obtained, by applying the estimated warping function on each of the original UV  parameters $\textbf{U}$.
The calculation of new UV parameters is necessary to map the faces to the corresponding regions in a newly warped texture map ($F_w$).

Finally, the downsampled RGB texture map $\mathbf{T}_{\downarrow s}$ is reconstructed from $F_w$. 
This downsampled texture map $\mathbf{T}_{\downarrow s}$ is overlaid on the surface of the mesh using the offset-adjusted UV parameters $\textbf{U'}$, yielding the mesh representation $\mathbf{M}(\mathbf{T}_{\downarrow s}, \mathbf{V}, \mathbf{E}, \mathbf{U'})$ as the final output.

\subsection{Rendering Loss and Optimization}
To optimize the parameters of the DNN, the differences between the rendered views generated from the original mesh and the output mesh are minimized.
At each iteration of training, a batch $\mathbf{P_K} = \{P_j\}_{j=1}^K$ of $K$ camera poses ($P_j$) focusing on the mesh are sampled randomly from a predefined range $\mathbb{S}$ of plausible and reasonable viewing conditions.
The range $\mathbb{S}$ is selected such that all reasonable viewing angles and distances are included within it. 
At each iteration of training, $K$ images are rendered from both the original mesh and the processed mesh with the cameras placed and oriented according to the set of sampled poses $\mathbf{P_K}$ as shown in \cref{eqn:imgs}
\begin{equation} \label{eqn:imgs}
    \begin{gathered}
        \mathbf{I_K^{orig}} = \{r(\mathbf{M}(\mathbf{T}, \mathbf{V}, \mathbf{E}, \mathbf{U}), P_j)\}_{P_j\in \mathbf{P_K}} \\
        \mathbf{I_K^{ds}} = \{r(\mathbf{M}(\mathbf{T_{\downarrow s}}, \mathbf{V}, \mathbf{E}, \mathbf{U'}), P_j)\}_{P_j\in \mathbf{P_K}}
    \end{gathered}
\end{equation}
where $r(\mathbf{M}(\cdot),P)$ is the rendering function that projects a textured mesh $\mathbf{M}(\cdot)$ to an RGB image from a viewing pose $P$, and $\mathbf{I_K^{orig}}$ and $\mathbf{I_K^{ds}}$ are the sets of images rendered from the original mesh and the mesh with the downsampled texture, respectively.
From these rendered images, a combination of 2D loss functions including L2 loss, SSIM  \cite{1284395} loss, and VGG-19  \cite{DBLP:journals/corr/SimonyanZ14a} based perceptual loss \cite{Johnson2016Perceptual} are computed. 
The losses of each pair of rendered images are summed to obtain the rendering loss for that iteration as shown in \cref{eqn:loss}.
\begin{equation} \label{eqn:loss}
    \begin{gathered}
        \text{Loss} = \sum_{j=1}^K[\alpha_1\mathcal{L}_{L2}(I_j^{orig}, I_j^{ds}) + \alpha_2\mathcal{L}_{SSIM}(I_j^{orig}, I_j^{ds}) + \alpha_3\mathcal{L}_{percep.}(I_j^{orig}, I_j^{ds})]
    \end{gathered}
\end{equation}
Given the availability of a plethora of differentiable rendering algorithms, one can be chosen that closely matches the target application.

The performance of the model is validated after every few iterations by following these steps:
\begin{enumerate}
    \item A set of $Q$ of poses $\mathbf{P_Q}$ is sampled uniformly from $\mathbb{S}$ and is predefined at the start of training.
    \item $Q$ images ($\mathbf{I_Q^{orig}}$) are rendered from camera poses in $\mathbf{P_Q}$ using the mesh with the original high-resolution texture.
    \item $Q$ images ($\mathbf{I_Q^{ds}}$) are rendered from camera poses in $\mathbf{P_Q}$ using the mesh with the downsampled texture.
    \item Metrics are calculated for each pair of images in $\mathbf{I_Q^{orig}}$ and $\mathbf{I_Q^{ds}}$ and averaged over the number of elements in the set
\end{enumerate}

\section{Experiments}
\label{sec:expt}

% \begin{table}[]
% \centering
% \resizebox{\columnwidth}{!}{%
% \begin{tabular}{|c|c|c|c|}
% \hline
% Scale &
%   \begin{tabular}[c]{@{}c@{}}GeoScaler \\ (Full)\end{tabular} &
%   w/o UVWarping &
%   \begin{tabular}[c]{@{}c@{}}w/o UVWarping \& \\ w/o GeoCoding\end{tabular} \\ \hline
% 4x &
%   \begin{tabular}[c]{@{}c@{}}35.05 /\\ 0.9564\end{tabular} &
%   \begin{tabular}[c]{@{}c@{}}34.45 /\\ 0.9515\end{tabular} &
%   \begin{tabular}[c]{@{}c@{}}34.18 /\\ 0.9486\end{tabular} \\ \hline
% 8x &
%   \begin{tabular}[c]{@{}c@{}}31.85 /\\ 0.9218\end{tabular} &
%   \begin{tabular}[c]{@{}c@{}}30.54\\ 0.8983\end{tabular} &
%   \begin{tabular}[c]{@{}c@{}}30.15 /\\ 0.8895\end{tabular} \\ \hline
% \end{tabular}%
% }
% \caption{\textbf{Ablation results of the main modules in the GeoScaler architecture.} Average PSNR(dB)/SSIM obtained on the 3DSet5 dataset before and after ablating the GeoCoding and UVWarp modules.}
% \label{tab:abla}
% \end{table}

% \begin{figure}
%     \centering
%     \includegraphics[width=8cm]{figures/memcmp.pdf}
%     \caption{
%     \textbf{Trade-off between rendering memory consumption and visual quality}. Textures produced by GeoScalar produce superior visual quality while consuming the same rendering memory compared to textures produced by bicubic downsampling.}
%     \label{fig:tradeoff}
% \end{figure}

\subsection{Implementation Details}
Our DNN model contains four residual blocks in the downsampling encoder and two graph residual blocks in the GeoCoding module. 
We chose $C_1=C_2=64$, bringing the total number of parameters in the network to 0.57M.
To compute the rendering loss, we used the differentiable renderer proposed in SoftRasterizer \cite{liu2019softras}, along with a Soft Phong Shader \cite{10.1145/15886.15897} whose implementations are available in PyTorch3D \cite{10.1145/3415263.3419160}.  
At each iteration, we rendered $K=4$ views when computing the loss. 
For the rendering step, we uniformly illuminated the scene with a white ambient light. 
The AdamW optimizer \cite{DBLP:conf/iclr/LoshchilovH19} was employed for the training process. 
We used a peak learning rate of $5e-3$ with a cosine annealing schedule \cite{DBLP:conf/iclr/LoshchilovH17} and a warm-up over 1000 iterations. 
We found that it is sufficient to train the DNN over 15000 iterations to achieve convergence.
When validating the model, we used $Q=60$ poses sampled uniformly over the range of camera poses.

\subsection{Datasets}
\label{subsec:dataset}
\begin{figure}
    \centering
    \includegraphics[width=12cm]{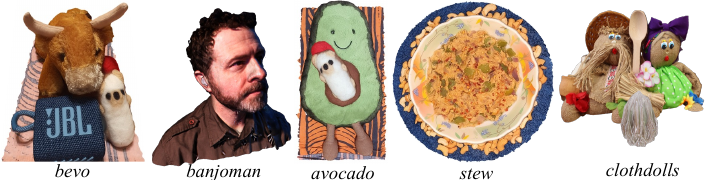}
    \caption{
    \textbf{The newly proposed 3DSet5 dataset}. Meshes in the 3DSet5 dataset are unprocessed and unrefined after reconstructing from multiple images of the objects and contain numerous geometric and textural irregularities.}
    \label{fig:dataset}
\end{figure}

% We tested GeoScaler on a wide variety of textured meshes selected from multiple datasets including Google's Scanned Object (GSO) dataset \cite{10.1109/ICRA46639.2022.9811809} and the Textured Mesh Quality Assessment (TMQA) dataset \cite{10.1145/3592786}.

We tested GeoScaler on a wide variety of textured meshes. 
We studied its efficacy on the entire Textured Mesh Quality Assessment Dataset (TMQA) \cite{10.1145/3592786} which consists of 55 meshes with diverse and varying geometry and color. 
This dataset contains meshes downloaded from SketchFab, and as such contains textured meshes generated using unknown yet varied 3D reconstruction algorithms and post-processing techniques. 
The curators of the dataset also portray its diversity in terms of color and geometric detail in \cite{10.1145/3592786}.
To facilitate experimentation, we decimate the meshes to contain a maximum of 20,000 faces and texture maps of 2048x2048 resolution.

We also studied the performance of GeoScaler on unprocessed and unrefined textured meshes which are directly generated from multiple images of real-world 3D objects, which contain geometric artifacts and irregularities.
To that end, we share a small dataset of five textured meshes that did not undergo any post-processing or refinement and present the performance of GeoScaler on it.
We dub this dataset 3DSet5 and its meshes are shown in \cref{fig:dataset}. 
The meshes in the 3DSet5 were generated from multiple input views (approx. 50) using Agisoft's Metashape application. 
The meshes in 3DSet5 contain approximately 10000 vertices and 20000 faces and have texture maps of size 2048x2048.
%We also demonstrate that downsampling the texture map using GeoScaler outputs higher quality renders compared to natively generating textures from Metashape at the reduced resolutions.
In the supplementary section, we analyze the performance of GeoScaler on the 3DSet5 dataset after the UV parametrizations of meshes were optimized with OptCuts\cite{10.1145/3272127.3275042}, which reduces the number of seams and warping artifacts in the texture maps.

These testing datasets were chosen such that the proposed method is evaluated on meshes of real-world objects that were generated using a diverse range of algorithms and that underwent varied post-processing techniques.

\subsection{Results and Ablation Study}

\begin{figure*}
    \centering
    \includegraphics[width=12cm]{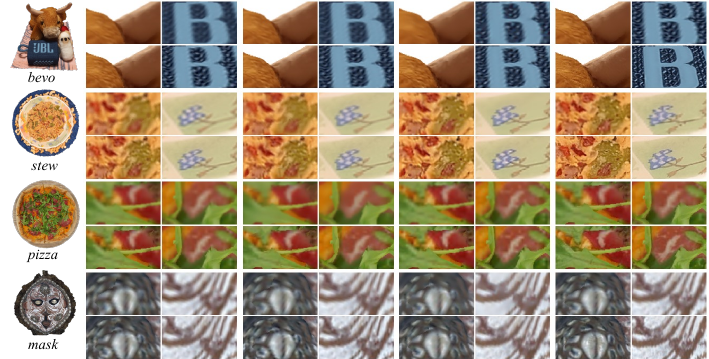}
    \caption{
    \textbf{Visual results ablating and comparing GeoScaler with existing methods for 4x downsampling}. The top row of each mesh shows the results using bicubic downsampling, Lanczos downsampling, "Perceptually based downsampling" and "GS Base" from left to right. The bottom row for each mesh shows the results using "GS Base+GCM", "GS Base+UVW", "GeoScaler Full" and the original from left to right.}
    \label{fig:subjective4x}
\end{figure*}

\begin{figure*}
    \centering
    \includegraphics[width=12cm]{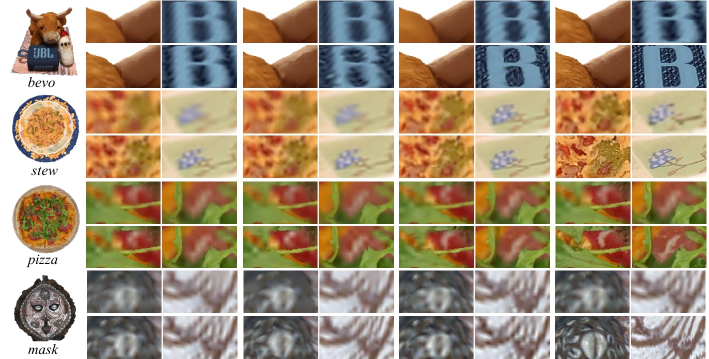}
    \caption{
    \textbf{Visual results ablating and comparing GeoScaler with existing methods for 8x downsampling}. The top row of each mesh shows the results using bicubic downsampling, Lanczos downsampling, "Perceptually based downsampling" and "GS Base" from left to right. The bottom row for each mesh shows the results using "GS Base+GCM", "GS Base+UVW", "GeoScaler Full" and the original from left to right.}
    \label{fig:subjective8x}
\end{figure*}

\begin{table}[]
\centering
\resizebox{\columnwidth}{!}{%
\begin{tabular}{|cc|c|c|c|c|c|c|c|}
\hline
\multicolumn{2}{|c|}{\textbf{Dataset}} &
  \textbf{Bicubic} &
  \textbf{Lanczos} &
  \textbf{PBD} &
  \textbf{GS Base} &
  \textbf{\begin{tabular}[c]{@{}c@{}}GS Base\\ +GCM\end{tabular}} &
  \textbf{\begin{tabular}[c]{@{}c@{}}GS Base\\ +UVW\end{tabular}} &
  \textbf{\begin{tabular}[c]{@{}c@{}}GeoScaler\\ Full\end{tabular}} \\ \hline
\multicolumn{1}{|c|}{\multirow{2}{*}{\textbf{TMQA}}}   & \textbf{PSNR (dB)} & 33.72  & 34.11  & 34.24  & 34.97  & 35.42  & 35.55  & 35.88  \\ \cline{2-9} 
\multicolumn{1}{|c|}{}                                 & \textbf{SSIM}      & 0.9264 & 0.9294 & 0.9286 & 0.9327 & 0.9428 & 0.9461 & 0.9524 \\ \hline
\multicolumn{1}{|c|}{\multirow{2}{*}{\textbf{3DSet5}}} & \textbf{PSNR (dB)} & 32.82  & 33.15  & 33.23  & 34.18  & 35.42  & 34.45  & 35.05  \\ \cline{2-9} 
\multicolumn{1}{|c|}{}                                 & \textbf{SSIM}      & 0.9350 & 0.9388 & 0.9421 & 0.9486 & 0.9428 & 0.9515 & 0.9564 \\ \hline
\end{tabular}%
}
\caption{\textbf{Quantitative results of 4x downscaling comparing GeoScaler and its ablations with existing methods.} GeoScaler outperforms traditional methods by a significant margin when applied to TMQA and 3DSet5 datasets.}
\label{tab:objective4x}
\end{table}

\vspace{-20px}

% \usepackage{multirow}
% \usepackage{graphicx}
\begin{table}[]
\centering
\resizebox{\columnwidth}{!}{%
\begin{tabular}{|cc|c|c|c|c|c|c|c|}
\hline
\multicolumn{2}{|c|}{\textbf{Dataset}} &
  \textbf{Bicubic} &
  \textbf{Lanczos} &
  \textbf{PBD} &
  \textbf{GS Base} &
  \textbf{\begin{tabular}[c]{@{}c@{}}GS Base\\ +GCM\end{tabular}} &
  \textbf{\begin{tabular}[c]{@{}c@{}}GS Base\\ +UVW\end{tabular}} &
  \textbf{\begin{tabular}[c]{@{}c@{}}GeoScaler\\ Full\end{tabular}} \\ \hline
\multicolumn{1}{|c|}{\multirow{2}{*}{\textbf{TMQA}}}   & \textbf{PSNR (dB)} & 30.48  & 30.88  & 31.02  & 31.74  & 32.41  & 32.42  & 33.18  \\ \cline{2-9} 
\multicolumn{1}{|c|}{}                                 & \textbf{SSIM}      & 0.8866 & 0.8935 & 0.8971 & 0.9066 & 0.9106 & 0.9115 & 0.9191 \\ \hline
\multicolumn{1}{|c|}{\multirow{2}{*}{\textbf{3DSet5}}} & \textbf{PSNR (dB)} & 28.63  & 29.34  & 29.53  & 30.15  & 32.41  & 30.54  & 31.85  \\ \cline{2-9} 
\multicolumn{1}{|c|}{}                                 & \textbf{SSIM}      & 0.8747 & 0.8779 & 0.8812 & 0.8996 & 0.9036 & 0.9047 & 0.9218 \\ \hline
\end{tabular}%
}
\caption{\textbf{Quantitative results of 8x downscaling comparing GeoScaler and its ablations with existing methods.} GeoScaler outperforms traditional methods by a significant margin when applied to TMQA and 3DSet5 datasets.}
\label{tab:objective8x}
\end{table}

We measured the quality gains provided by GeoScaler in terms of peak signal-to-noise ratio (PSNR) and the Structural Similarity Index Metric (SSIM) as shown in \cref{tab:objective4x} \cref{tab:objective8x}.
GeoScaler was able to significantly outperform existing methods including bicubic downsampling, Lanczos downsampling, and "Perceptually Based Downsampling" (PBD) proposed by \cite{Oztireli15Perceptual} by improving downsampling quality at both 4x and 8x scale factors.
These tables also demonstrate the performance of GeoScaler after the proposed modules were ablated. 
The GeoScaler Base (GS Base) model is obtained by ablating both the GeoCoding module and the UVWarper Module. 
Its performance verifies the impact of using rendering loss and performing optimization based on the rendered space rather than the texture's UV plane.
We add the GeoCoding module to GS Base to obtain the "GS Base + GCM" model whose performance gain over GS Base validates the GeoCoding module.
Similarly, "GS Base + UVW" obtained by adding only UVWarping to GS Base tests this module independently.

The improvement in downsampling quality is also evident for scale factors of 4x and 8x in the visual results shown in \cref{fig:subjective4x} and \cref{fig:subjective8x} respectively. 
The meshes \textit{mask} and \textit{pizza} shown in the figure are from the Textured Mesh Quality Assessment dataset \cite{10.1145/3592786}, while \textit{bevo} and \textit{stew} are from 3DSet5.
It can be noticed that GeoScaler displays significantly superior retention of details and fine structures contained in the original texture. 
GeoScaler also prevents artifacts at seams on textures, as shown in 8x downsampling results of \textit{bevo} (horizontal seam artifact in the left columns are reduced when Geocoding is applied).

\begin{figure}
    \centering
    \includegraphics[width=12cm]{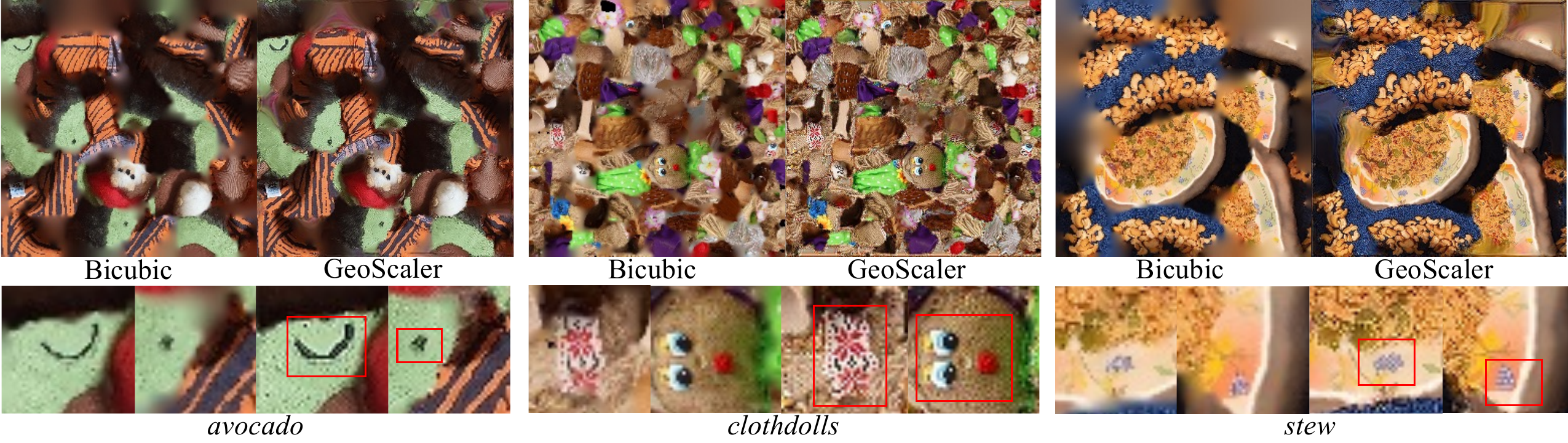}
    \caption{
    \textbf{Optimization of UV mapping by UVWarper}. Highly detailed regions are assigned larger areas in texture maps produced by GeoScaler. Regions of interest marked by red boxes demonstrate the improved space allocation.}
    \label{fig:textures}
\end{figure}

\begin{figure}
    \centering
    \includegraphics[width=12cm]{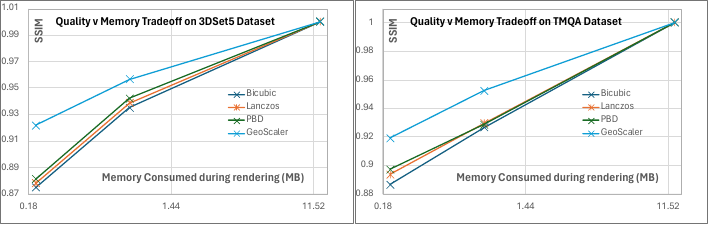}
    \caption{
    \textbf{Trade-off between rendering memory consumption and visual quality}. Textures produced by GeoScalar produce superior visual quality while consuming same rendering memory compared to textures produced using existing methods}
    \label{fig:tradeoff}
\end{figure}

The effect of the UVWarper module is also visualized in \cref{fig:textures} where the texture map itself is displayed.
The figure shows the reassignment of areas in the texture map based on the salience of RGB details in the texture map.
In the supplementary material, we also show that textures generated by GeoScaler provide perceptually better results than textures generated natively at lower resolution using MetaShape over the 3DSet5 dataset.

Using GeoScaler, the trade-off between memory usage while rendering and visual quality is greatly improved, with noticeably improved visual quality of the rendered contents even though much lower rendering run-time memory was required.
The average reduction in SSIM is plotted against memory consumed while rendering in \cref{fig:tradeoff} to highlight this aspect.
The memory consumed is equal to the total number of pixels in the texture, multiplied by its bit-depth and by the number of color channels.

\section{Conclusions}
\label{sec:conclusions}
We presented a method of downsampling texture maps that is superior to existing methods using a unified representation of texture maps and the geometries of 3D meshes that they are painted on.
This framework is also applicable for downsampling other feature maps associated with meshes, such as roughness and specularity.
GeoScaler-based downsampling of textures is intended to enable the deployment of compact graphical assets in applications for devices having low computation and memory budgets such as smartphones and VR headsets while maintaining high visual quality. 
It also facilitates remastering high-quality graphical assets for video games on devices with lower rendering capabilities.

While direct potential negative impacts of this work are not anticipated, the method has a few limitations.
Firstly, this method cannot be directly used to generate texture maps from images captured in the wild as it relies on the presence of a high-resolution texture map.
Consequently, this method is also unable to rectify and would replicate any errors or artifacts present in the original texture.
Secondly, while a per-mesh optimization framework enables content-adaptive downsampling and significantly enhanced quality, it requires much more processing time than the existing methods that are practically instantaneous.
These limitations pave the way for interesting future work on this topic.

\bibliographystyle{splncs04}
\bibliography{main}

\section{Supplementary Material}
\label{sec:suppl}

\subsection{Ablating the Skip Residual across GeoCoding}
\label{subsec:s1}
\begin{figure}
    \centering
    \includegraphics[width=12cm]{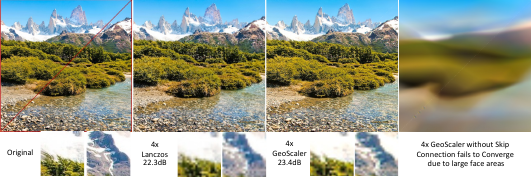}
    \caption{\textbf{Ablating the Skip Residual:} When the area of faces in texture maps is large, the skip connection allows bypassing the GeoCoding module and mitigates the loss in reconstruction quality}
    \label{fig:skipabla}
\end{figure}

As explained in \cref{subsec:gcm}, a skip residual across the GeoCoding module is introduced to mitigate the drop in performance when the vertices are extremely sparse, and the sampled vectors at the vertices fail to retain sufficient information.
In other words, when the area of faces on the texture map is large, sampling features $F_e$ at vertex locations would result in a large drop in the amount of information retained that is necessary for the reconstruction of the texture map. 
In these cases, the skip connection allows an alternate route for feature flow that bypasses the GeoCoding module as a fallback.
This can be demonstrated using a simple example where GeoScaler is applied on a square mesh having 2 large textured faces (see \cref{fig:skipabla}).

Without the skip connection, the model fails to converge, and can only capture low-frequency details, whereas the original model performs normally. 
When faces are not too large, the GeoCoding module remains effective as evidenced by the ablations shown in \cref{tab:objective4x} and \cref{tab:objective8x}.

\subsection{Evaluating GeoScaler on Meshes with Optimized UV Mappings}
\label{subsec:s2}
While demonstrating the performance of GeoScaler on the entire TMQA dataset and the 3DSet5 dataset covers meshes built and post-processed using a wide range of algorithms, the performance of GeoScaler when the UV mapping of the mesh is "optimal" remains to be seen.
An "optimal" mapping reduces the seams in the texture map and attempts to keep the ratio of areas of triangles in 3D and areas of corresponding triangles in the UV plane as uniform as possible.
Algorithms like OptCuts \cite{10.1145/3272127.3275042} and AutoCuts \cite{10.1145/3130800.3130845} are capable of performing the optimization.
To demonstrate the performance of GeoScaler on meshes with optimized UVs we first generate fresh UV mapping of the meshes in 3DSet5 using OptCuts \cite{10.1145/3272127.3275042}, bake the original texture to the new maps, and then apply our method.
The results shown in \cref{tab:tabopt} indicate that GeoScaler continues to improve downsampling performance for meshes with optimized textures.

% \usepackage{graphicx}
% \usepackage{graphicx}
\begin{table}[]
\centering
\resizebox{\columnwidth}{!}{%
\begin{tabular}{|c|c|c|c|c|c|c|}
\hline
\textbf{Scale} & \textbf{Bicubic} & \textbf{Lanczos} & \textbf{GS Base} & \textbf{GS   Base+GCM} & \textbf{GS   Base+UVW} & \textbf{GeoScaler (Full)} \\ \hline
\textbf{4x}    & 33.80 / 0.9490   & 33.91 / 0.9500   & 34.33 / 0.9531   & 34.39 / 0.9537         & 34.55 / 0.9554         & 35.28 / 0.9610            \\ \hline
\textbf{8x}    & 29.17 / 0.9056   & 29.23 / 0.9068   & 29.51 / 0.9142   & 29.66 / 0.9149         & 30.25 / 0.9168         & 31.04 / 0.9222            \\ \hline
\end{tabular}%
}
\caption{\textbf{Performance of GeoScaler on meshes with optimized UV mappings.} The two numbers in each cell indicate the PSNR(dB) and SSIM.}
\label{tab:tabopt}
\end{table}

\subsection{Comparing with Generating Textures in Lower Resolution}
We also compare the quality of textures downsampled by GeoScaler on the 3DSet5 dataset with the textures generated natively at 4x and 8x lower resolutions. 
Note that the meshes and textures in the 3DSet5 dataset were reconstructed using a proprietary tool MetaShape (see \cref{subsec:dataset}).
We use the same application and pipeline for generating textures at 4x and 8x lower resolutions.
The results in \cref{tab:meta} suggest that downsampling textures after reconstructing 3D scenes at higher resolutions can provide higher-quality renders than generating low-resolution textures directly on 3Dset5.

\begin{table}
\centering
\caption{\textbf{Quantitative results comparing GeoScaler with generating textures at lower resolutions natively}}
\label{tab:meta}
\begin{tblr}{
  cells = {c},
  cell{2}{1} = {r=2}{},
  cell{4}{1} = {r=2}{},
  vlines,
  hline{1-2,4,6} = {-}{},
  hline{3,5} = {2-6}{},
}
\textbf{Scale} & \textbf{Metric}    & \textbf{MetaShape} & \textbf{Bicubic} & \textbf{Lanczos} & {\textbf{GeoScaler}\\\textbf{ Full}} \\
\textbf{4x }   & \textbf{PSNR (dB)} & 32.34              & 33.72            & 34.11            & 35.88                                \\
               & \textbf{SSIM}      & 0.9022             & 0.9264           & 0.9294           & 0.9524                               \\
\textbf{8x}    & \textbf{PSNR (dB)} & 26.56              & 28.63            & 29.34            & 31.85                                \\
               & \textbf{SSIM}      & 0.8510             & 0.8747           & 0.8779           & 0.9218                               
\end{tblr}
\end{table}

We suspect MetaShape uses bilinear resampling internally to subsample the images fed during the reconstructing process to generate texture maps leading to poor quality.
This leads to interesting future research where per-scene optimization methods such as ours can be used for generating higher-quality texture maps at lower resolutions during the 3D scene reconstruction process.

\subsection{More Results}
\begin{figure}
    \centering
    \includegraphics[width=12cm]{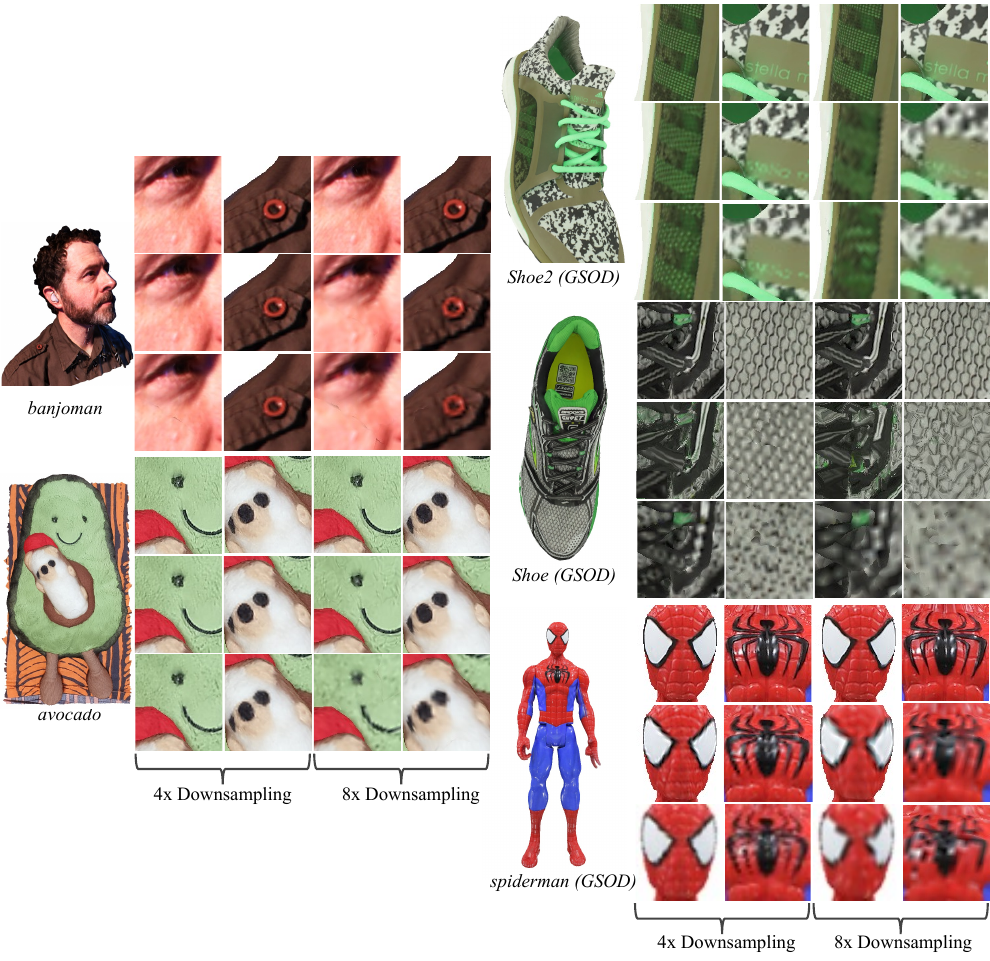}
    \caption{\textbf{More Results on remaining 3DSet5 meshes and a few meshes from Google's Scanned Objects Dataset}For each mesh result, the top row is the Ground Truth texture, the middle row shows results from GeoScaler, and the bottom row shows results using Bicubic}
    \label{fig:moreres}
\end{figure}

The results on \textit{banjoman} and \textit{avocado} from 3DSet5 are also shown in \cref{fig:moreres}.
Additionally, we also show results on a few meshes sampled from Google's Scanned Object dataset in \cref{fig:moreres}.

\textbf{All the meshes for which results are shown are included in the zipped file along with the supplementary materials.}

\end{document}